\documentclass{article}

 \usepackage[preprint]{creativeai}
 \usepackage{amsmath,amssymb,amsfonts} 
 \usepackage{bm} 
 \usepackage{graphicx}  

\usepackage[utf8]{inputenc} 
\usepackage[T1]{fontenc}    
\usepackage{hyperref}       
\usepackage{url}            
\usepackage{booktabs}       
\usepackage{amsfonts}       
\usepackage{nicefrac}       
\usepackage{microtype}      
\usepackage{xcolor}         
\usepackage{float} 

\title{TraitSpaces: Towards Interpretable Visual Creativity for Human–AI Co-Creation}

\author{%
  Prerna Luthra\\
  Evam Labs\\
  Singapore \\
  \texttt{pl2243@nyu.edu}
}

\begin{document}

\maketitle

\begin{abstract}
We introduce a psychologically grounded and artist-informed framework for modeling visual creativity across four domains—Inner, Outer, Imaginative, and Moral Worlds. Drawing on interviews with practicing artists and theories from psychology, we define 12 traits that capture affective, symbolic, cultural, and ethical dimensions of creativity. Using 20k artworks from the SemArt dataset, we annotate images with GPT-4.1 using detailed, theory-aligned prompts, and evaluate the learnability of these traits from CLIP image embeddings. Traits such as Environmental Dialogicity and Redemptive Arc are predicted with high reliability ($R^2 \approx 0.64 - 0.68$), while others like Memory Imprint remain challenging, highlighting the limits of purely visual encoding. Beyond technical metrics, we visualize a “creativity trait-space” and illustrate how it can support interpretable, trait-aware co-creation - e.g., sliding along a Redemptive Arc axis to explore works of adversity and renewal. By linking cultural-aesthetic insights with computational modeling, our work aims not to reduce creativity to numbers, but to offer shared language and interpretable tools for artists, researchers, and AI systems to collaborate meaningfully.

\end{abstract}

\section{Introduction}
Creativity assessment in AI has largely relied on narrow proxies—fluency/originality scores from psychometric traditions [Torrance, 1974; Kaufman \& Beghetto, 2009] or technical image metrics like FID [Heusel et al., 2017] and CLIPScore [Hessel et al., 2021]—that overlook affective, symbolic, cultural, and ethical dimensions. While broader theories frame creativity as emerging from person–process–product–context interactions [Amabile, 1983; Csikszentmihályi, 1996; Boden, 2004], operationalizing such frameworks for vision-language models remains challenging due to tensions between psychological richness and computational tractability.

To bridge this gap, we synthesize psychological theory with artist interviews into a four-world framework (Inner, Outer, Imaginative, Moral). Its 12 traits (e.g., Emotional Intensity, Environmental Dialogicity) balance conceptual depth with partial computability, enabling application across creative domains. We focus on visual art—where these dimensions manifest concretely in composition using 20k SemArt images [García \& Vogiatzis, 2018].

Using GPT-4.1 as a scalable annotation proxy [Tan et al., 2024], we score artworks per trait, then regress against CLIP embeddings [Radford et al., 2021] to: 
 
(1) Quantify visual encodability of creative dimensions, 

(2) Derive interpretable trait directions for semantic control (e.g., "increase Redemptive Arc"),  

(3) Propose Human AI co-creation tool framework based on trait directions.  

This establishes the foundation for trait-aware human–AI collaboration beyond novelty-centric evaluation. 
While our framework introduces computable dimensions, it is not intended to reduce creativity to numbers. Instead, it offers a shared language to connect artistic, cultural, and psychological insights with interpretable computational tools.

\section{4-World Creativity Framework}
We developed this framework through semi-structured interviews with practicing artists and writers about their creative process, and then synthesized those insights with foundational psychological and creativity theories. Participants often described creativity as a form of exploration, of inner experience, the external world, imaginative possibility, and moral concern, an inductive theme that, together with prior theory, motivated our four 'world' framework.
The resulting taxonomy organizes twelve traits into four domains—the \emph{Inner}, \emph{Outer}, \emph{Imaginative}, and \emph{Moral} worlds—that reflect complementary facets of creative practice.

\paragraph{Inner World.}
Human creativity also involves what we term the \emph{Inner World}—a cluster of introspective dimensions grounded in affect, memory, and symbolic cognition. We operationalize this as three core components: \textbf{Emotional Intensity}, \textbf{Memory Imprint}, and \textbf{Personal Symbolism}.

\emph{Emotional Intensity} captures the raw immediacy and introspective depth of affective experience. Rooted in psychodynamic theory [Freud, 1908], intense emotion is sublimated into symbolic form, while narrative psychology emphasizes emotion as central to life-story construction [McAdams, 2001]. \emph{Memory Imprint} reflects the autobiographical embedding of lived experience—drawing on sensory detail, flashbacks, and self-referential cues that ground the work in personal history [Conway \& Pleydell-Pearce, 2000; McAdams, 2001]. Finally, \emph{Personal Symbolism} involves a creator’s idiosyncratic system of metaphors, archetypes, and surreal logic, anchored in depth psychology [Jung, 1964]. These dimensions align with foundational theories of internal ideation in the Geneplore model [Finke et al., 1992], pointing to psychologically rich and authentic aspects of creative practice.

\paragraph{Outer World.}
Beyond internal ideation, creative work is deeply embedded in the \emph{Outer World}—shaped by culture, environment, and social discourse. We define this domain through three interrelated constructs: \textbf{Cultural Situatedness}, \textbf{Environmental Dialogicity}, and \textbf{Social Reflexivity}.

\emph{Cultural Situatedness} refers to how grounded a work is in a specific cultural, historical, or geographic context. Drawing from sociocultural theory [Vygotsky, 1978; Glăveanu, 2014], it reflects how creators engage with shared cultural tools, traditions, and domains. \emph{Environmental Dialogicity} captures how the physical or natural environment participates in meaning-making, aligning with ecocritical frameworks that treat landscapes as active co-agents in narrative [Buell, 2005; Bate, 2000]. Finally, \emph{Social Reflexivity} involves critical engagement with audience and power structures, grounded in dialogic and performative theory [Bakhtin, 1981; Butler, 1997], whereby creative works both reflect and challenge the norms of their social milieu. Together, these dimensions frame creativity as interaction across systems of meaning.

\paragraph{Imaginative World.}
Creativity within the \emph{Imaginative World} reflects a symbolic, surreal, and playful cognitive space characterized by three interrelated dimensions: \textbf{Surreal Divergence}, \textbf{Symbolic Density}, and \textbf{Playful Subversion}.

\emph{Surreal Divergence} captures the blending of reality with dreamlike or fantastical elements that subvert ordinary logic and perception. Rooted in psychoanalytic theory [Freud, 1919; Jung, 1964], this dimension evokes visionary, subconscious-driven experiences through otherworldly imagery that defies everyday expectations. \emph{Symbolic Density} denotes multi-layered, richly woven symbolism, where motifs or events carry multiple, often ambiguous meanings, inviting diverse interpretations and deep engagement [Fauconnier \& Turner, 2002; Eco, 1976]. Finally, \emph{Playful Subversion} embodies the freedom to bend or break conventions—mixing genres and employing irony, absurdity, and novelty to surprise and engage the audience—resonating with postmodern aesthetics of disruption and hybridity [Hutcheon, 1989]. Collectively, these dimensions define a richly imaginative space where symbolic complexity and playful originality converge.

\paragraph{Moral World.}
Creativity in the \emph{Moral World} engages ethical, communal, and transformative dimensions, situating works as arenas for moral reflection and social change. We conceptualize this domain through \textbf{Ethical Provocation}, \textbf{Collective Resonance}, and \textbf{Redemptive Arc}.

\emph{Ethical Provocation} reflects a work’s capacity to confront audiences with complex moral dilemmas or injustices, provoking reflection without simple resolutions, consistent with Nussbaum’s notion of moral imagination [Nussbaum, 1990]. \emph{Collective Resonance} denotes the articulation and amplification of communal experiences—particularly those of marginalized groups—fostering empathy and solidarity [Polletta, 2006; hooks, 1994]. Finally, the \emph{Redemptive Arc} captures trajectories where adversity catalyzes healing, transformation, or hope, reflecting McAdams’s redemptive self-stories and Campbell’s universal hero’s journey [McAdams, 2006; Campbell, 1949]. Together, these facets frame creativity as ethical engagement, communal meaning-making, and social renewal.

\section{Method}

\paragraph{Dataset:}We use the SemArt dataset, a multimodal corpus of Western fine-art paintings (approximately 21,000 images) [García \& Vogiatzis, 2018]. The training, test, and validation sets contain about 19,000, 1,000, and 1,000 images, respectively, as provided by the dataset.

\paragraph{GPT-4.1 trait scoring:} We operationalize the 12-trait rubric by passing detailed prompts. These prompts provide full trait definitions and a 0-4 integer rubric (Appendix B). For each image--trait pair, we query GPT-4.1 (\texttt{gpt-4-turbo-2024-04-09}) with temperature $=0$ and \texttt{top\_p} $=1$, instructing it to return only an integer in $\{0,1,2,3,4\}$

\paragraph{CLIP embeddings:} Each image is encoded with CLIP ViT-B/32; we use the 512-D image embedding vector [Radford et al., 2021].

\paragraph{Ridge probe (interpretable trait axes):} Let $\mathbf{X}_{\mathrm{train}}\in\mathbb{R}^{N\times 512}$ be the CLIP ViT-B/32 image embeddings for the training images, \emph{after} per-sample $L_2$ normalization and mean-centering with the train mean $\boldsymbol{\mu}_{\mathrm{train}}$ (the same $\boldsymbol{\mu}_{\mathrm{train}}$ is reused for validation and test to avoid leakage). For a given trait $T$, let $\mathbf{y}_{T,\mathrm{train}}\in\{0,1,2,3,4\}^N$ be the GPT-4.1–assigned integer scores for those images (the targets). We fit Ridge (no intercept, fixed regularization) by
\[
\min_{\mathbf{w}_T}\ \bigl\lVert \mathbf{X}_{\mathrm{train}}\mathbf{w}_T - \mathbf{y}_{T,\mathrm{train}}\bigr\rVert_2^2
\;+\; \lambda \lVert \mathbf{w}_T\rVert_2^2,
\]
yielding a \emph{trait direction} $\mathbf{w}_T$. We then learn a 1-D linear calibration on train projections $s_{\mathrm{train}}=\mathbf{X}_{\mathrm{train}}\mathbf{w}_T$,
\[
\hat{\mathbf{y}}_{T,\mathrm{train}} \;=\; a_T\, s_{\mathrm{train}} + b_T,
\]
and apply it to test: $\hat{\mathbf{y}}_{T,\mathrm{test}}=\mathrm{clip}\!\left(a_T(\mathbf{X}_{\mathrm{test}}\mathbf{w}_T)+b_T,\,[0,4]\right)$, where $\mathbf{X}_{\mathrm{test}}$ uses the same $L_2$ normalization and train-mean centering.

\paragraph{XGBoost regressors (accuracy-oriented):} Separately for each trait $T$, we train an XGBoost regressor on the \emph{raw} CLIP ViT-B/32 embeddings from the train split (no $L_2$ or mean-centering), using the validation split for early stopping (30 rounds). Test predictions are clipped to $[0,4]$. Note that we used standard gradient-boosting settings (e.g., max\_depth~6, learning\_rate~0.05, n\_estimators up to 700, subsample/colsample\_bytree~$<1$).

\paragraph{Evaluation and visualization:} Spearman’s $\rho$ (rank correlation) between image embedding predictions $\hat{\mathbf{y}}_{T,\mathrm{test}}$ and GPT\mbox{-}4.1 scores $\mathbf{y}_{T,\mathrm{test}}$; $R^2$ and Mean Absolute Error (MAE) computed on the same pair.
For visualization, we compute UMAP (cosine metric) on normalized embeddings and plot trait directions as finite-difference displacements.

\section{Results and Discussion}

\begingroup
\setlength{\textfloatsep}{\baselineskip}   
\setlength{\abovecaptionskip}{\baselineskip}  
\setlength{\belowcaptionskip}{\baselineskip}  

\begin{table}[t]
\centering
\footnotesize
\setlength{\tabcolsep}{6pt}
\caption{semArt test results using CLIP (ViT-B/32, 512-d) embeddings with XGBoost. Rows ordered by $R^2$ (XGB) descending on the test split ($N{=}1067$).}
\begin{tabular}{l r r r}
\toprule
Trait & $R^2$ (XGB) & $\rho$ (XGB) & MAE (XGB) \\
\midrule
Environmental Dialogicity & 0.677 & 0.810 & 0.355 \\
Redemptive Arc            & 0.639 & 0.801 & 0.455 \\
Ethical Provocation       & 0.635 & 0.801 & 0.444 \\
Emotional Intensity       & 0.609 & 0.760 & 0.250 \\
Collective Resonance      & 0.607 & 0.786 & 0.385 \\
Symbolic Density          & 0.590 & 0.771 & 0.257 \\
Personal Symbolism        & 0.555 & 0.745 & 0.427 \\
Social Reflexivity        & 0.519 & 0.719 & 0.387 \\
Cultural Situatedness     & 0.514 & 0.701 & 0.247 \\
Surreal Divergence        & 0.497 & 0.691 & 0.460 \\
Playful Subversion        & 0.325 & 0.550 & 0.299 \\
Memory Imprint            & 0.287 & 0.557 & 0.321 \\
\bottomrule
\end{tabular}
\end{table}
\endgroup

\paragraph{Overall performance:} On the SemArt test split, several traits are predicted from CLIP (ViT-B/32, 512-d) embeddings with high fidelity. XGBoost (in Table 1) reaches Spearman $\rho \approx 0.76$--$0.81$ and $R^2 \approx 0.59$--$0.68$ for \emph{Environmental Dialogicity, Ethical Provocation, Redemptive Arc, Symbolic Density, Collective Resonance}, and \emph{Emotional Intensity}. Ridge regression, used to recover interpretable trait directions, yields nearly the same rank ordering and only slightly lower $R^2$ ($\Delta R^2 \le 0.031$), suggesting that the mapping from embeddings to trait scores is approximately linear. Refer to results in Appendix A. Mid-tier traits cluster around $R^2 \approx 0.51$--$0.56$ (\emph{Personal Symbolism, Social Reflexivity, Cultural Situatedness}), with \emph{Surreal Divergence} slightly lower ($R^2 \approx 0.50$). The hardest dimensions are \emph{Playful Subversion} and \emph{Memory Imprint} ($R^2 \approx 0.33$ and $0.29$, respectively). Aggregating by the four ``worlds'' in the rubric, mean XGBoost $R^2$ is highest for the \textbf{Moral} world ($\approx 0.63$), followed by the \textbf{Outer} world ($\approx 0.57$), and lower for the \textbf{Inner} and \textbf{Imaginative} worlds ($\approx 0.48$ and $\approx 0.47$).

\paragraph{Trait semantics and difficulty:} To interpret these performance differences, we revisit the rubric’s definitions. \emph{Environmental Dialogicity} evaluates whether the environment functions as an active presence; \emph{Ethical Provocation} seeks depicted dilemmas; \emph{Redemptive Arc} captures adversity$\rightarrow$renewal trajectories; \emph{Symbolic Density} emphasizes layered motifs; \emph{Collective Resonance} requires communal perspectives; \emph{Emotional Intensity} combines visceral and introspective elements. These manifest through visually grounded cues (color, objects, composition) well represented in CLIP [Kwon \& Ye, 2022; Patashnik et al., 2021; Wu \& Maji, 2022]. Conversely, \emph{Memory Imprint} (autobiographical specificity) and \emph{Playful Subversion} (irony/genre-mixing) require extra-visual context, aligning with CLIP's weaknesses in spatial/compositional reasoning [Kamath et al., 2023; Thrush et al., 2022]. \emph{Social Reflexivity}'s ambiguity---evident in high human inter-annotator variance for similar constructs [Achlioptas et al., 2021]---explains its mid-tier predictability.
The difficulty of predicting traits such as Memory Imprint and Playful Subversion highlights that creativity is not wholly reducible to visual features; it also depends on cultural context, autobiographical depth, and lived experience that remain beyond current models.

\begin{figure}
    \centering
    \includegraphics[width=0.98\textwidth]{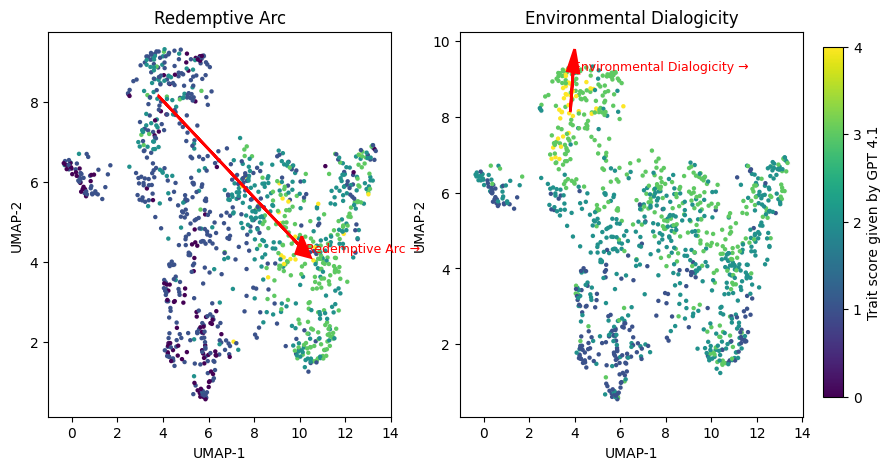}
    \caption{UMAP projection of trait space, with Ridge direction arrows.}
    \label{fig:umap}
\end{figure}

\paragraph{Geometry of the trait space:} As shown in Figure~1, we visualize a UMAP of CLIP (ViT-B/32) image embeddings from the SemArt test set, color-coded by GPT-4.1--derived trait scores. Overlaid arrows indicate the Ridge-derived directions for each trait, computed as finite differences in normalized embedding space. For the best-predicted traits---\emph{Environmental Dialogicity} and \emph{Redemptive Arc}---the color gradient increases monotonically along the corresponding arrow. This alignment provides qualitative support for interpreting each trait as an approximate axis in the embedding space, consistent with the rubric’s semantics.

\paragraph{Towards Human--AI Co-Creation:}
Our findings suggest CLIP-based trait directions enable interpretable controls, while XGBoost provides reliable assessment meters. Together, they can help artists and designers explore creativity dimensions, articulate intent, and iterate efficiently. For example: A designer drags an \textit{Environmental Dialogicity} slider from low$\rightarrow$high, shifting a UMAP corpus view from indoor portraits to weather-dominated landscapes.  Clicking a high-intensity cluster surfaces exemplar images; bookmarking these as references, the system suggests \textit{compositionally similar but trait-enhanced variants} via CLIP-guided generation for rapid iteration.  

We allocate control using empirical predictability thresholds:  
\begin{itemize}
  \item \textbf{Direct control} ($R^2 \geq 0.55$, $\rho \geq 0.70$): \textit{Environmental Dialogicity, Ethical Provocation, Redemptive Arc, Symbolic Density, Collective Resonance, Emotional Intensity} (and borderline \textit{Personal Symbolism}). Traits are steered via Ridge sliders with XGBoost real-time scoring.  
  \item \textbf{Assisted control}: \textit{Social Reflexivity, Cultural Situatedness, Surreal Divergence}. System proposes candidates; humans refine/approve.  
  \item \textbf{Context-driven control}: \textit{Memory Imprint, Playful Subversion}. Humans specify intent via text/references; AI assists only in coarse filtering.  
\end{itemize}
\noindent Note that the present thresholds rely on GPT-derived labels; formal co-design and human-subject studies are future work.

\paragraph{Limitations:}
While our framework advances interpretable creativity modeling, several limitations remain. First, our experiments rely on the SemArt dataset, which is confined to Western fine-art traditions and thus carries cultural and stylistic biases that may not generalize to other artistic contexts. Second, the trait definitions are grounded in specific psychological and theoretical traditions and operationalized via prompts to GPT-4.1; changes in prompt phrasing or model behavior could lead to variations in the resulting scores. Third, our evaluation depends entirely on GPT-4.1–derived labels, as no large-scale, diverse human rating benchmark currently exists to validate the robustness of trait scoring across audiences. Addressing this gap will require coordinated efforts from both the research and broader artist communities to design and conduct large-scale annotation studies. Our approach should thus be viewed as a complement to, rather than a replacement for, human and cultural perspectives on creativity. Future work should integrate diverse voices, contexts, and multimodal forms to ensure that the richness of creative practice is not oversimplified.

\section{Call for Collaboration}
As this work represents an early step toward building a richer, interpretable framework for human–AI collaboration in creativity, \textbf{mentors, collaborators, and allies are invited} to help expand, critique, and co-develop these ideas. This research is exploratory and still at an early stage. While developed during my role at Evam, I am eager to continue building it further. If you are interested, you can reach out to me at \textbf{pl2243@nyu.edu}

\section{Acknowledgment}
We would like to thank everyone at Evam Labs and Team Poiro for their invaluable support—whether through debugging code, brainstorming ideas, or connecting us with artists and writers whose insights were instrumental in shaping this framework. We are especially grateful to the artists and writers who generously shared their time and creative perspectives during our conversations. In Appendix C, we feature selected works from artists who kindly consented to share their creations, in the hope of giving their work the visibility and appreciation it deserves.

.

\section*{References}

{\small

[1] Torrance, E.P. (1974) \textit{Torrance Tests of Creative Thinking}. Personnel Press, Bensenville, IL.

[2] Kaufman, J.C. \& Beghetto, R.A. (2009) Beyond Big and Little: The Four C Model of Creativity. \textit{Review of General Psychology}, 13(1), 1--12.

[3] Heusel, M., Ramsauer, H., Unterthiner, T., Nessler, B. \& Hochreiter, S. (2017) GANs trained by a two time-scale update rule converge to a local Nash equilibrium. \textit{Advances in Neural Information Processing Systems (NeurIPS)}.

[4] Hessel, J., Lee, L. \& Choi, Y. (2021) CLIPScore: A reference-free evaluation metric for image captioning. \textit{Proc. EMNLP}.

[5] Amabile, T.M. (1983) The social psychology of creativity: A componential conceptualization. \textit{Journal of Personality and Social Psychology}, 45(2), 357--376.

[6] Csikszentmihalyi, M. (1996) \textit{Creativity: Flow and the Psychology of Discovery and Invention}. Harper Perennial, New York.

[7] Boden, M.A. (2004) \textit{The Creative Mind: Myths and Mechanisms} (2nd ed.). Routledge, London.

[8] García, S. \& Vogiatzis, G. (2018) SemArt: A dataset for semantic art understanding. \textit{ECCV Workshops (Computer Vision for Art Analysis - CVAA)}.

[9] Tan, Z., Li, D., Wang, S., Beigi, A., Jiang, B., Bhattacharjee, A., Karami, M., Li, J., Cheng, L. \& Liu, H. (2024) Large Language Models for Data Annotation \textit{and Synthesis}: A Survey. \textit{arXiv preprint} arXiv:2402.13446. 

[10] Radford, A., Kim, J.W., Hallacy, C., et al. (2021) Learning transferable visual models from natural language supervision. \textit{Proc. ICML}.

[11] Freud, S. (1908) Creative Writers and Day-Dreaming. \textit{The Standard Edition of the Complete Psychological Works of Sigmund Freud}, Volume IX (1906--1908).

[12] McAdams, D.P. (2001) The psychology of life stories. \textit{Review of General Psychology}, 5(2), 100--122.

[13] Conway, M.A. \& Pleydell-Pearce, C.W. (2000) The construction of autobiographical memories in the self-memory system. \textit{Psychological Review}, 107(2), 261--288.

[14] Jung, C.G. (1964) \textit{Man and His Symbols}. Doubleday.

[15] Finke, R.A., Ward, T.B. \& Smith, S.M. (1992) \textit{Creative Cognition: Theory, Research, and Applications}. MIT Press.

[16] Vygotsky, L.S. (1978) \textit{Mind in Society: The Development of Higher Psychological Processes}. Harvard University Press.

[17] Glăveanu, V.P. (2014) The social nature of creativity: Conceptual and empirical arguments. \textit{Psychology of Aesthetics, Creativity, and the Arts}, 8(3), 277--279.

[18] Buell, L. (2005) \textit{The Environmental Imagination: Thoreau, Nature Writing, and the Formation of American Culture}. Harvard University Press.

[19] Bate, J. (2000) \textit{The Song of the Earth}. Harvard University Press.

[20] Bakhtin, M.M. (1981) \textit{The Dialogic Imagination: Four Essays}. University of Texas Press.

[21] Butler, J. (1997) \textit{Excitable Speech: A Politics of the Performative}. Routledge.

[22] Freud, S. (1919) The Uncanny. \textit{The Standard Edition of the Complete Psychological Works of Sigmund Freud}, Volume XVII (1917--1919).

[23] Fauconnier, G. \& Turner, M. (2002) \textit{The Way We Think: Conceptual Blending and the Mind's Hidden Complexities}. Basic Books.

[24] Eco, U. (1976) \textit{A Theory of Semiotics}. Indiana University Press.

[25] Hutcheon, L. (1989) \textit{The Politics of Postmodernism}. Routledge.

[26] Nussbaum, M.C. (1990) \textit{Love's Knowledge: Essays on Philosophy and Literature}. Oxford University Press.

[27] Polletta, F. (2006) \textit{It Was Like a Fever: Storytelling in Protest and Politics}. University of Chicago Press.

[28] hooks, b. (1994) \textit{Teaching to Transgress: Education as the Practice of Freedom}. Routledge.

[29] McAdams, D.P. (2006) The redemptive self: Generativity and the stories Americans live by. \textit{Research in Human Development}, 3(2--3), 81--100.

[30] Campbell, J. (1949) \textit{The Hero with a Thousand Faces}. Princeton University Press.

[31] Chen, T. \& Guestrin, C. (2016) XGBoost: A scalable tree boosting system. \textit{Proc. KDD}, 785--794.

[32] McInnes, L., Healy, J., \& Melville, J. (2020). UMAP: Uniform Manifold Approximation and Projection for Dimension Reduction. Journal of Open Source Software, 5(52), 861.

[33] Kwon, Y. \& Ye, J.C. (2022) Towards better understanding of CLIP: a review of evaluation methodologies. \textit{arXiv preprint arXiv:2207.10597}.

[34] Patashnik, O., Wu, Z., Shechtman, E., Cohen-Or, D., \& Lischinski, D. (2021). StyleCLIP: Text-Driven Manipulation of StyleGAN Imagery.  \textit{Proc. ICCV}.

[35] Wu, H. \& Maji, S. (2022) Vision-Language Pretrained Models: A Survey. \textit{IEEE Transactions on Pattern Analysis and Machine Intelligence}.

[36] Kamath, A., Denton, E., Farhadi, A., et al. (2023) METER: Measuring and evaluating text-to-image retrieval models. \textit{arXiv preprint arXiv:2301.11310}.

[37] Thrush, E., Kale, D., Singhal, K., et al. (2022) Evaluating compositional reasoning in vision-language models. \textit{arXiv preprint arXiv:2207.00012}.

[38] Achlioptas, P., Schluter, N., Wonka, P., \& Guibas, L. (2021) Learning representations for artistic style. \textit{IEEE Transactions on Pattern Analysis and Machine Intelligence}.

}

\clearpage
\appendix

\section{Ridge Regression Metrics}
\begin{table}[H]
\centering
\footnotesize
\setlength{\tabcolsep}{6pt}
\caption{SemArt test results using CLIP (ViT-B/32, 512-d) embeddings with \textbf{Ridge}. Rows ordered by $R^2$ (Ridge) descending on the test split. $\Delta R^2$ is (XGB $-$ Ridge).}
\begin{tabular}{l r r r r}
\toprule
Trait & $R^2$ (Ridge) & $\rho$ (Ridge) & MAE (Ridge) & $\Delta R^2$ \\
\midrule
Environmental Dialogicity & 0.673 & 0.812 & 0.364 &  0.004 \\
Ethical Provocation       & 0.649 & 0.811 & 0.450 & -0.014 \\
Redemptive Arc            & 0.640 & 0.804 & 0.463 & -0.001 \\
Collective Resonance      & 0.607 & 0.788 & 0.395 &  0.000 \\
Symbolic Density          & 0.598 & 0.784 & 0.273 & -0.008 \\
Emotional Intensity       & 0.578 & 0.752 & 0.287 &  0.031 \\
Personal Symbolism        & 0.540 & 0.740 & 0.433 &  0.015 \\
Social Reflexivity        & 0.521 & 0.720 & 0.394 & -0.002 \\
Surreal Divergence        & 0.494 & 0.686 & 0.478 &  0.003 \\
Cultural Situatedness     & 0.490 & 0.696 & 0.265 &  0.024 \\
Playful Subversion        & 0.325 & 0.569 & 0.324 &  0.000 \\
Memory Imprint            & 0.265 & 0.543 & 0.342 &  0.022 \\
\bottomrule
\end{tabular}
\end{table}

\clearpage

\section{Prompt Definitions given to GPT-4.1}

\subsection*{I. Inner World}
\begin{description}
\item[\textbf{1. Emotional Intensity}] Assess the immediacy and authenticity of emotion within the work. Does it convey visceral feelings through bold gestures, charged language, or powerful imagery? At the same time, does it invite deep self-reflection via internal monologues, existential questions, or other introspective elements? A work high in Emotional Intensity combines raw affect with profound introspection, making the audience feel strong emotions and ponder the creator’s inner thoughts.
\item[\textbf{2. Memory Imprint}] Consider how the work incorporates personal, autobiographical memory. Does it include sensory details, symbolic objects, or narrative flashbacks that evoke specific lived moments from the creator’s past? Such elements give the work a sense of temporal depth, anchoring it in the creator’s own history and leaving a lasting memory trace in the narrative or imagery.
\item[\textbf{3. Personal Symbolism}] Identify any unique symbolic system or dream-like logic present in the work that reflects the creator’s inner world. Does the creator use recurring motifs, archetypal images, or personal metaphors that form a cohesive, surreal narrative language unique to them? This dimension highlights idiosyncratic symbolism—a one-of-a-kind mythology or logic the creator has built into the piece to represent their personal psyche.
\end{description}

\subsection*{II. Outer World}
\begin{description}
\item[\textbf{4. Cultural Situatedness}] Examine how deeply the work is rooted in a specific cultural, geographic, or historical context. Does it draw on local traditions, dialects, landscapes, or historical references to provide a palpable sense of place and heritage? High Cultural Situatedness means the piece feels grounded in a particular community or environment, giving the audience a strong sense of where and when the story or artwork belongs.
\item[\textbf{5. Environmental Dialogicity}] Evaluate how the work engages with its physical or natural setting. Does it treat the environment not just as a backdrop but as an active presence or character that influences the narrative? For example, are places or natural forces portrayed as shaping human experiences or interacting with characters? In a highly environmentally dialogic piece, the landscape itself participates in meaning-making, engaging in a kind of dialogue with the human elements of the work.
\item[\textbf{6. Social Reflexivity}] Determine how the work acknowledges its audience and reflects on the social context. Does it speak directly to the viewer or reader (for example, through rhetorical questions or by breaking the fourth wall)? Does it critically examine social norms, power structures, or injustices within its content? A work high in Social Reflexivity invites the audience to help construct its meaning—it engages the viewer as a participant—while provoking awareness of collective issues in society.
\end{description}

\subsection*{III. Imaginative World}
\begin{description}
\item[\textbf{7. Surreal Divergence}] Look for any blending of reality with dreamlike or fantastical elements. Does the work subvert ordinary logic and perception, introducing bizarre scenarios or otherworldly imagery that feel like a dream or vision? Surreal Divergence is characterized by a distortion of reality that taps into the subconscious or altered states. High presence of this dimension means the piece creates a visionary, almost dream-world experience that defies everyday expectations.
\item[\textbf{8. Symbolic Density}] Analyze how rich and layered the symbolism is in the work. Do individual images, events, or motifs carry multiple meanings and invite varied interpretations? A piece with high Symbolic Density packs a lot of meaning into its symbols or metaphors, rewarding close reading or viewing with new associations. Such a work weaves many possible interpretations into a single element, creating depth and complexity in its narrative or imagery.
\item[\textbf{9. Playful Subversion}] Identify elements of playfulness and radical originality in the work’s concept or form. Does it bend rules or mix genres, perhaps using irony, absurdity, wordplay, or non-linear storytelling techniques? Also consider the uniqueness of its ideas: does the piece introduce unprecedented concepts or metaphors that defy convention? High Playful Subversion is evident when a work feels whimsical or experimental in style while also showing innovative originality in content, surprising the audience with something truly novel.
\end{description}

\subsection*{IV. Moral World}
\begin{description}
\item[\textbf{10. Ethical Provocation}] Gauge how strongly the work provokes moral questioning or confronts the audience with ethical dilemmas. Does it present conflicting values, dilemmas, or injustices that cause discomfort or force reflection? Rather than offering clear-cut resolutions, a work high in Ethical Provocation will challenge the audience’s sense of right and wrong, compelling them to examine their own moral assumptions and feelings of urgency around the issues raised.
\item[\textbf{11. Collective Resonance}] Consider whether the work speaks to the experiences or identity of a larger group or community, especially one that is marginalized or underrepresented. Does it move beyond a single individual’s perspective to give voice to a collective narrative or shared emotional landscape? A work with strong Collective Resonance fosters empathy and solidarity; it resonates with a group’s struggles or hopes and allows audiences (within or outside that group) to connect with those communal experiences on an emotional level.
\item[\textbf{12. Redemptive Arc}] Observe whether the work contains a trajectory of transformation, healing, or hope. Does the narrative (or imagery) move from adversity toward reconciliation, justice, or spiritual renewal? A strong Redemptive Arc means that despite any hardship or darkness depicted, the piece ultimately offers a sense of resolution or uplift. It traces a path where characters or themes undergo meaningful change—providing the audience with feelings of catharsis, redemption, or hope by the end of the work.
\end{description}

\clearpage
\section{Artist Creations}
\begin{figure}[H]
    \centering
    \includegraphics[width=0.5\textwidth]{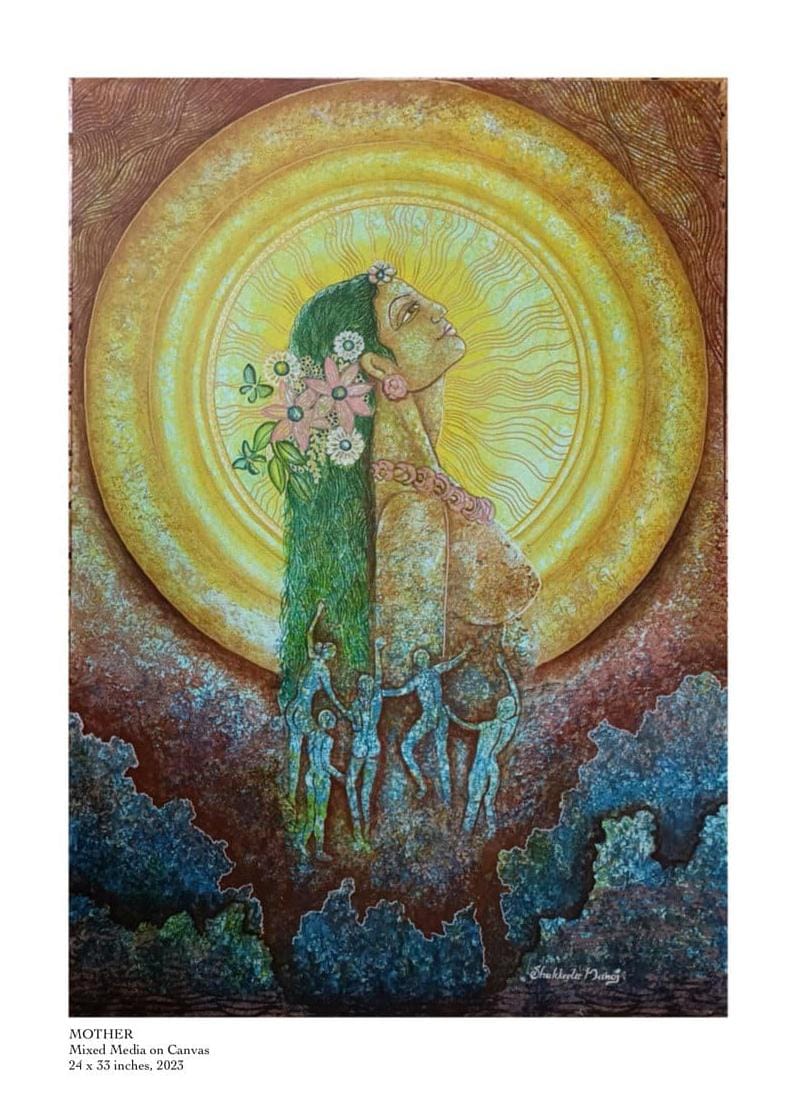}
    \caption{Mother by Shakkeela Manoj}
    \label{fig:artist_top}
\end{figure}

\vfill 

\begin{figure}[b]
    \centering
    \includegraphics[width=0.5\textwidth]{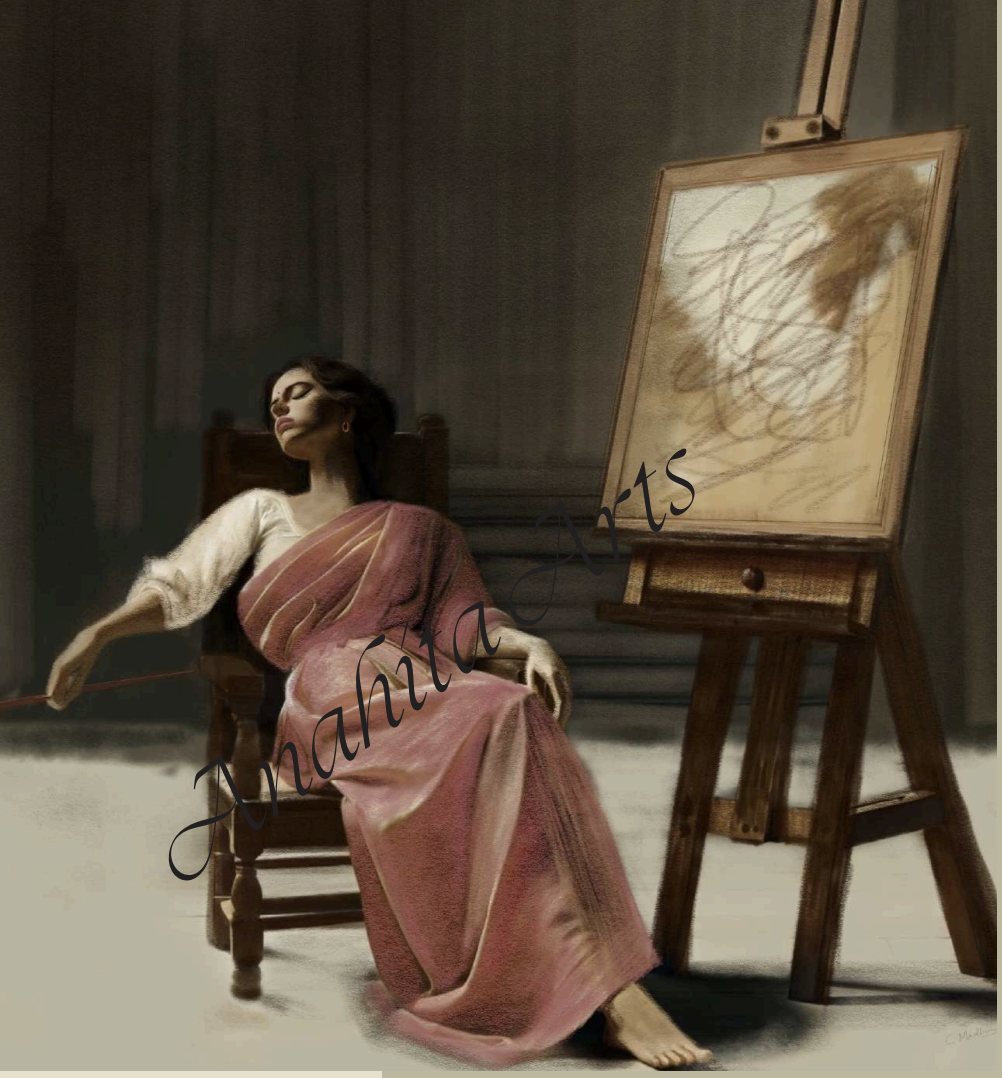}
    \caption{What the Canvas knows by Madhurima Cumsali}
    \label{fig:artist_bottom}
\end{figure}


\end{document}